\documentclass[12pt]{iopart}
\usepackage{iopams}
\usepackage[nosort]{cite}
\usepackage[bookmarks=false]{hyperref}
\hypersetup{colorlinks=true,citecolor=blue,linkcolor=blue,urlcolor=blue,pdfstartview=FitH,bookmarksopen=true}
\begin{document}
\title[]{A proof of the Kochen-Specker theorem can always be converted to a state-independent noncontextuality inequality}
\author{Xiao-Dong Yu$^1$, Yan-Qing Guo$^2$, and D. M. Tong$^1$}
\address{$^1$Department of Physics, Shandong University, Jinan 250100, China}
\address{$^2$Department of Physics, Dalian Maritime University, Dalian 116026, China}
\ead{\mailto{tdm@sdu.edu.cn}}

\begin{abstract}
Quantum contextuality is one of the fundamental notions in quantum mechanics. Proofs of the Kochen-Specker theorem and noncontextuality inequalities are two means for revealing the contextuality phenomenon in quantum mechanics. It has been found that some proofs of the Kochen-Specker theorem, such as those based on rays, can be converted to a state-independent noncontextuality inequality, but it remains open whether it is true in general, i.e., whether any proof of the Kochen-Specker theorem can always be converted to a noncontextuality inequality. In this paper, we address this issue. We prove that all kinds of proofs of the Kochen-Specker theorem, based on rays or any other observables, can always be converted to state-independent noncontextuality inequalities. Besides, our constructive proof also provides a general approach for deriving a state-independent noncontextuality inequality from a proof of the Kochen-Specker theorem.
\end{abstract}



\submitto{\NJP}


\section{Introduction}

Quantum contextuality \cite{Kochen.Specker1967} as a natural generalization of Bell nonlocality \cite{Bell1966} is one of the fundamental notions in quantum mechanics, and has drawn a lot of interest recently. Proofs of the Kochen-Specker (KS) theorem and noncontextuality inequalities are two different methods for revealing the contextuality phenomenon in quantum mechanics. Both of them have been widely used in many fields of quantum mechanics, including quantum computation \cite{Raussendorf2013,Howard.etal2014,Delfosse.etal2015} and quantum information \cite{Horodecki.etal2010,Cabello.etal2011,Nagali.etal2012,Um.etal2013,Guehne.etal2014,Canas.etal2014}.

KS theorem was found from the inconsistency of the noncontextual hidden variable (NCHV) model with quantum mechanics.  The NCHV model \cite{Kochen.Specker1967,Bell1966} consists of two basic assumptions: that every observable $A$ has a definite value $v(A)$ at all time, and that $v(A)$ does not depend on whether $A$ is measured alone or together with $B$ or $C$ if $A$ is compatible with $B$ and $C$. The first assumption is at the core of the hidden variable theory while the latter is the exhibition of noncontextuality.
An observable corresponds to a Hermitian operator in quantum mechanics, and two observables being compatible in the NCHV model is corresponding to the operators being commutative in quantum mechanics. A measurement context in the NCHV model, defined as a set of mutually compatible observables, is corresponding to a set of mutually commutative operators in quantum mechanics. To examine the inconsistency of the NCHV model with quantum mechanics, Kochen and Specker assumed that the algebraic structure of compatible observables in quantum mechanics is also preserved in the NCHV model, which leads to the sum rule and product rule \cite{Fine.Teller1978,Redhead1987}. The sum rule means that if $A$ and $B$ are two compatible observables and $C=A+B$, then $v(C)=v(A)+v(B)$. Similarly, the product rule means that if $A$ and $B$ are two compatible observables and $C=A\cdot B$, then $v(C)=v(A)\cdot v(B)$.  It was found that there always exists a finite set of observables for any $n$-dimensional Hilbert space with $n\ge 3$, such that all elements in the set cannot simultaneously have values satisfying the sum rule and product rule. This finding is usually called the KS theorem, which shows that the NCHV model is not compatible with quantum mechanics.

A proof of the KS theorem is accomplished by finding an observable set in which no value assignment of the observables can satisfy the sum rule and product rule, i.e., it inevitably  leads to a logical contradiction if the value assignments of the observables are required to satisfy the sum rule and product rule. Such sets of the observables that can lead to a logical contradiction between the value assignments and the algebraic structure are not unique, and each of them provides a proof of the KS theorem. Early proofs of the KS theorem are all based on the rays, rank-$1$ projectors, in the Hilbert space. The original proof of the KS theorem proposed by Kochen and Specker involves 117 rays. The size of the ray sets is reduced step by step \cite{Peres1991,Bub1996,Conway.Kochen,Kernaghan1994,Cabello.etal1996} and the simplest proofs based on rays are with 31 rays for a $3$-dimensional Hilbert space \cite{Conway.Kochen} and 18 rays for a $4$-dimensional Hilbert space \cite{Cabello.etal1996}. Generalization of proofs of the KS theorem started from the works of Mermin and Peres \cite{Mermin1990,Peres1991}, who found that the number of observables involved in a proof of the KS theorem can be significantly reduced if general observables are used, instead of only using rays. The Mermin-Peres square proposed by them involves only $9$ observables in a $4$-dimensional Hilbert space. Since then, a great number of proofs of the KS theorem based on general observables have been proposed \cite{Pagonis.etal1991,Kernaghan.Peres1995,Cabello1999,Cabello2001,Cerf.etal2002,Planat2012,Waegell.Aravind2012,Waegell.Aravind2013,Ruuge2012,Tang.etal2013,Lisonek.etal2014b,Toh2013a,Toh2013b}.

Noncontextuality inequalities are an alternative tool for demonstrating the inconsistency between the NCHV model and quantum mechanics. A noncontextuality inequality is an expression which is fulfilled for all value assignments of observables in the NCHV model but is violated for some states in quantum mechanics. A noncontextuality inequality is said to be state-independent if it is violated by all quantum states. Proofs of the KS theorem and noncontextuality inequalities are two ways of showing the inconsistency between the NCHV model and quantum mechanics. Compared with proofs of the KS theorem, noncontextuality inequalities are more experiment-friendly, and therefore have drawn a lot of interest recently \cite{Klyachko.etal2008,Cabello.etal2008,Cabello2008,Badziag.etal2009,Guehne.etal2010,Liang.etal2011,Cabello2011,Yu.Oh2012,Abramsky.Brandenburger2011,Bengtsson.etal2012,Kleinmann.etal2012,Kurzynski.Kaszlikowski2012,Kurzynski.etal2012,Chaves.Fritz2012,Amselem.etal2012,Araujo.etal2013,Yu.Tong2014,Ramanathan.Horodecki2014,Lisonek.etal2014a,Winter2014,Su.etal2015,Acin.etal2015}.

Cabello may be the first who noted that some proofs of the KS theorem can be used to construct state-independent noncontextuality inequalities, and proposed the first state-independent noncontextuality inequality \cite{Cabello2008}. Soon after, Badzi\c{a}g \textit{et al} showed that every proof of the KS theorem based on rays can be converted to a state-independent noncontextuality inequality \cite{Badziag.etal2009}. Indeed, for a proof of the KS theorem based on the ray set $\mathcal{S}=\{P_1,P_2, \dots, P_\mu\}$ in an $n$-dimensional Hilbert space, one can construct the function $F=\sum_{\alpha=1}^N\left(\sum_{i=1}^nA_{k_i^\alpha}-\prod_{i=1}^n(1+A_{k_i^\alpha})\right)$, where $A_{k_i^\alpha}=1-2P_{k_i^\alpha}$, and $\{P_{k_1^\alpha},P_{k_2^\alpha},\dots,P_{k_n^\alpha}\}\in S$ is the $\alpha$-th basis of the space, $\alpha=1,2,\dots,N$ with $N$ being the total number of the bases. If $\mathcal{S}$ satisfies the condition that every pair of orthogonal rays belongs to some $\{P_{k_1^\alpha},P_{k_2^\alpha},\dots,P_{k_n^\alpha}\}$, then one can immediately obtain the noncontextuality inequality for the NCHV model, $\langle F\rangle\le N(n-2)-2$, which is violated for all quantum states since $\langle F\rangle=N(n-2)$ in quantum mechanics. If $\mathcal{S}$ does not satisfy this condition, it should be enlarged by adding new rays to satisfy the condition and an inequality can be derived from the enlarged set. Most recently, the result of Badzi\c{a}g \textit{et al} was improved by Yu and Tong in Ref. \cite{Yu.Tong2014}, where the authors put forward an alternative method, which need not enlarge the given ray set $\mathcal{S}$ and can give a simpler inequality from a proof of the KS theorem based on rays.

These previous results show that some proofs of the KS theorem, such as those based on rays, can be converted to state-independent noncontextuality inequalities. However, it remains open whether such correspondence is true in general, i.e., whether any proof of the KS theorem, not limited to those based on rays, can always be converted to a state-independent noncontextuality inequality. This is a fundamental issue related to the relation between proofs of the KS theorem and noncontextuality inequalities. In this paper, we address this issue. Our results show that every proof of the KS theorem, based on rays or other observables, can always be converted to a state-independent noncontextuality inequality. Besides, our constructive proof also provides a simple approach for deriving a state-independent noncontextuality inequality from a general proof of the KS theorem. Proofs of the KS theorem based on general observables are usually simpler than those only based on rays, and therefore the inequalities derived from them are expected to be simpler and more useful.

This paper is organized as follows. In section 2, we demonstrate the existence of a noncontextuality inequality in any proof of the KS theorem and put forward a general approach to derive the inequality. In sections 3 and 4, as examples of showing its application, we apply our approach to proofs of the KS theorem based on rays and proofs of the KS theorem based on parity arguments, respectively, and derive the corresponding noncontextuality inequalities. Section 4 is the conclusion and remarks.

\section{A general approach of converting a proof of the KS theorem to a noncontextuality inequality}

We first specify some notation and relations. $\mathcal{S}$ is used to denote a set of observables which carries out a proof of the KS theorem, $\mathcal{S}=\{A_1, A_2, \dots, A_\mu\}$, where the observables $A_i$ may be rays or other observables.
A measurement context is defined as a subset of $\mathcal{S}$, denoted as $\mathcal{S}_\alpha$, in which all observables are mutually compatible, where the subscript $\alpha $ is used to label different measurement contexts, $\alpha =1,2,\dots, L$. The $\alpha$-th measurement context can be written as $\mathcal{S}_\alpha=\{A_{k^\alpha_1}, A_{k^\alpha_2}, \dots, A_{k^\alpha_m}\}$, where the subscripts $1\le k^\alpha_1<k^\alpha_2<\cdots<k^\alpha_m\le\mu$ and $m=m(\alpha)$ is the number of observables in the $\alpha$-th measurement context. Note that $m$ in the subscript of $k^\alpha_m$ depends on $\alpha$, i.e., the number of observables in different measurement contexts may be different. We use $\mathbb{R}[A_{k^\alpha_1}, A_{k^\alpha_2}, \dots, A_{k^\alpha_m}]$ to denote the set of all polynomials of $A_{k^\alpha_1}, A_{k^\alpha_2}, \dots, A_{k^\alpha_m}$, with real coefficients. $\mathcal{I}_\alpha$ is a subset of $\mathbb{R}[A_{k^\alpha_1}, A_{k^\alpha_2}, \dots, A_{k^\alpha_m}]$, defined as $\mathcal{I}_\alpha=\{r\in\mathbb{R}[A_{k^\alpha_1}, A_{k^\alpha_2}, \dots, A_{k^\alpha_m}]| r(\hat{A}_{k^\alpha_1}, \hat{A}_{k^\alpha_2}, \dots, \hat{A}_{k^\alpha_m})=0\}$, where $\hat{A_i}$ is the operator of observable $A_i$.
The terminology, a `value assignment' $v$ to $\mathcal{S}$ means that a value, denoted as $v(A_i)$, is assigned to each observable $A_i$ in $\mathcal{S}$,
where $v(A_i)$ is an eigenvalue of the operator $\hat{A}_i$.
If a value assignment $v$ satisfies the sum rule and product rule, then $r|_v\equiv r(v(A_{k^\alpha_1}), v(A_{k^\alpha_2}), \dots, v(A_{k^\alpha_m}))=0$ for $r\in\mathcal{I}_\alpha$, $\alpha=1,2,\dots,L$, since the observables $A_{k^\alpha_1}, A_{k^\alpha_2}, \dots, A_{k^\alpha_m}$ are mutually compatible.

We now demonstrate the existence of a noncontextuality inequality in a proof of the KS theorem expressed by $\mathcal{S}$. The fact that the observable set $\mathcal{S}$ provides a proof of the KS theorem means that no value assignment $v$ can simultaneously fulfil all the equations in $\{r|_v=0|r\in\cup_{\alpha=1}^L\mathcal{I}_\alpha\}$, which is a infinite set since each $\mathcal{I}_\alpha$ is infinite. Note that $\mathcal{I}_\alpha$ is an ideal of the polynomial ring $\mathbb{R}[A_{k^\alpha_1}, A_{k^\alpha_2}, \dots, A_{k^\alpha_m}]$. According to Hilbert's basis theorem, $\mathbb{R}[A_{k^\alpha_1}, A_{k^\alpha_2}, \dots, A_{k^\alpha_m}]$ is a Noetherian ring, and therefore $\mathcal{I}_\alpha$ can be finitely generated. For each $\mathcal{I}_\alpha$, we can always find a finite subset of $\mathcal{I}_\alpha$, $\{r_1^\alpha,r_2^\alpha,\dots,r^\alpha_{I_\alpha}\}$, such that all $r\in\mathcal{I}_\alpha$ can be written as $r=\sum_{i=1}^{I_\alpha}f_i r_i^\alpha$ with $f_i\in\mathbb{R}[A_{k^\alpha_1}, A_{k^\alpha_2},\dots, A_{k^\alpha_m}]$. This implies that $r|_v=0$ for all $r\in\mathcal{I}_\alpha$ if and only if $r_i^\alpha|_v=0$ for $i=1,2,\dots,I_\alpha$. Hence, the observable set $\mathcal{S}$ provides a proof of the KS theorem if and only if no value assignment $v$ can simultaneously fulfil the equations, $\{r_i^\alpha|_v=0|i=1,2,\dots,I_\alpha, \alpha=1,2,\dots,L\}$, which include a finite number of polynomials $r_i^\alpha$.
By using these polynomials, we can define a function, $F=-\sum_{\alpha=1}^L\sum_{i=1}^{I_\alpha} \left(r_i^\alpha(A_{k^\alpha_1},A_{k^\alpha_2},\dots,A_{k^\alpha_m})\right)^2$. It is easy to verify that $\langle F\rangle<0$ plays the role of a state-independent noncontextuality inequality. In fact, since at least one $r_i^\alpha|_v$ is nonzero for any value assignment $v$ and therefore $F|_v=-\sum_{\alpha=1}^L\sum_{i=1}^{I_\alpha} (r_i^\alpha|_v)^2<0$, the average value $\langle F\rangle$ must be less than zero. On the other hand, by definition, there is $r_i^\alpha(\hat{A}_{k^\alpha_1},\hat{A}_{k^\alpha_2},\dots,\hat{A}_{k^\alpha_m})=0$. We have $\langle F\rangle=\mathrm{Tr}(F\rho)=-\sum_{\alpha=1}^L\sum_{i=1}^{I_\alpha}\mathrm{Tr}(({r_i^\alpha})^2\rho)=0$ for all quantum states $\rho$, which means that all quantum states violate the inequality $\langle F\rangle<0$.

With the help of the above demonstration, we may now develop a general approach of converting a proof of the KS theorem to a state-independent noncontextuality inequality. The above discussions show that the key is to find a set of observable polynomials, $\{r_1,r_2,\dots,r_N\}$, which satisfies the two conditions: that $r_i$ is a polynomial of some measurement context $\mathcal{S}_\alpha=\{A_{k^\alpha_1}, A_{k^\alpha_2}, \dots, A_{k^\alpha_m}\}$ satisfying that $r_i (\hat{A}_{k^\alpha_1}, \hat{A}_{k^\alpha_2}, \dots, \hat{A}_{k^\alpha_m})=0$, and that for any value assignment $v$, at least one of $\{r_1|_v,r_2|_v,\dots,r_N|_v\}$ is nonzero. For convenience, hereafter, we call such a set of observable polynomials that satisfy the two conditions as a \textit{complete} set of polynomials. Clearly, for a given set of observables $\mathcal{S}$, there may be many different \textit{complete} sets of polynomials and the number $N$ for different sets may be different too. We only need to find one of them to construct a noncontextuality inequality in $\mathcal{S}$. Although it is difficult to prove whether an observable set provides a proof of the KS theorem, it is trivial to find a \textit{complete} set of polynomials from the observable set that has been proved to be a proof of the KS theorem. Indeed, if a set of observables is a proof of the KS theorem, it means that no value assignment to the set can satisfy the sum rule and product rule. This is verified either by finding, for each value assignment, a polynomial of compatible observables that is equal to zero in quantum mechanics but nonzero for the value assignment in the NCHV model, or by finding several polynomials of compatible observables that are equal to zero in quantum mechanics but lead to a logical contradiction if they are required to be zero in the NCHV model. In each of the two cases, the set of the polynomials used in the proof is just what we want to find. They comprise a \textit{complete} set of polynomials, $\{r_1,r_2,\dots,r_N\}$, and can be used to construct a noncontextuality inequality, $\langle F\rangle<0$ with $F=-\sum_{i=1}^N r_i^2$. If we would like to obtain an inequality being with an explicit upper bound, we may normalize each polynomial by $r_i/\sqrt{c_i}$, where $c_i=\min_{r_i|_v\ne 0}\left(r_i|_v\right)^2$ is a constant. In this case, there is $r_i^2|_v\ge 1$ if $r_i|_v=0$ is violated. By using these normalized polynomials $r_i/\sqrt{c_i}$ to take the place of $r_i$, the inequality can be expressed as $\langle F\rangle\le-1$. Besides, the state-independent inequality obtained can be simplified by reducing $A_i^k$ ($k\ge d$) with $(A_i-a_1)(A_i-a_2)\cdots(A_i-a_d)=0$, where $a_1,a_2\dots,a_d$ are the $d$ different eigenvalues of $A_i$.

So far, we have demonstrated the existence of a noncontextuality inequality in any proof of the KS theorem and have shown how to convert a proof of the KS theorem to a noncontextuality inequality. To summarize our approach briefly, one may derive a state-independent noncontextuality inequality from a proof of the KS theorem expressed by a set of observables $\mathcal{S}=\{A_1,A_2,\dots,A_\mu\}$ via the following three steps.
\\$1)$ Find a \textit{complete} set of normalized polynomials, denoted as
\begin{equation}
\mathcal{P}_c=\{r_1,r_2,\dots,r_N\}.
 \label{PC}
\end{equation}
This can be done by first finding a \textit{complete} set of polynomials and then normalizing them by multiplying each by a suitable constant.
\\$2)$ Define a function of observables by using the normalized polynomials,
\begin{equation}
F=-\sum_{i=1}^N r_i^2,
\label{eq:defF}
\end{equation}
and simplify $F=F(A_1,A_2,\dots,A_\mu)$ by using the relation $(A_i-a_1)(A_i-a_2)\cdots(A_i-a_d)=0$.
\\$3)$ Write out the expression,
\begin{equation}
\langle F(A_1,A_2,\dots,A_\mu)\rangle<-1,
\label{eq:ineq}
\end{equation}
which is a state-independent noncontextuality inequality, i.e., violated by all the quantum states.

It is worth noting that in the above discussions, the polynomials $r_i$ have been assumed to be real for simplification. Although this assumption is applicable for almost all known proofs of the KS theorem, there are a few proofs of the KS theorem in which complex polynomials are involved \cite{Cerf.etal2002,Tang.etal2013}. Our approach is also applicable to these cases. When complex polynomials are involved in a proof of the KS theorem, the only modification is that equation (\ref{eq:defF}) is replaced by
\begin{equation}
  F=-\sum_{i=1}^N r_i^\dagger r_i.
  \label{eq:defFc}
\end{equation}

In the following sections, we will take two well-known kinds of proofs of the KS theorem, proofs based on rays and proofs based on parity arguments, as examples to illustrate the approach.

\section{Application to proof of the KS theorem based on rays}

Let $\mathcal{S}$ be a set of rays that provides a proof of the KS theorem, $\mathcal{S}=\{P_1,P_2,\dots,P_\mu\}$, where $P_i$ are rays in an $n$-dimensional Hilbert space. A ray set provides a proof of the KS theorem if and only if no value assignment $v$ can simultaneously satisfy the KS rules: (i) $v(P_i)v(P_j)=0$ if $P_i$ and $P_j$ are orthogonal; and (ii) $\sum_{i=1}^nv(P_{k_i^\alpha})=1$, if $\{P_{k_i^\alpha}|i=1,2,\dots,n\}$ forms an orthogonal basis for the $n$-dimensional Hilbert space. In other words, at least one of these equations is invalid for each value assignment.

We now derive a state-independent noncontextuality inequality from the set $\mathcal{S}$ by using our approach described in section 2. First, $\{P_iP_j~|~\hat{P}_i\hat{P}_j=0,~P_i,P_j\in\mathcal{S}\}\cup\{\sum_{i=1}^nP_{k_i^\alpha}-1~|~\sum_{i=1}^n\hat{P}_{k_i^\alpha}=I_n,~P_{k_i^\alpha}\in\mathcal{S}\}$ comprise a \textit{complete} set of polynomials \footnote{If the ray set is denoted by a graph, in which each vertex corresponds to a ray and two vertices corresponding to two orthogonal rays are connected by an edge, then these polynomials just correspond to the edges and n-vertex cliques in the graph.}, and each of the polynomials is already normalized since $v(P_i)v(P_j)=1$ if $v(P_i)v(P_j)\ne 0$ and $\sum_{i=1}^nv(P_{k_i^\alpha})-1=-1$ or $\geq 1$ if $\sum_{i=1}^n v(P_{k_i^\alpha})-1\ne 0$. Therefore, we may take this set of polynomials as $\mathcal{P}_c$, a \textit{complete} set of normalized polynomials defined in equation (\ref{PC}). Second, substituting all the polynomials for $r_i$ in equation (\ref{eq:defF}), and using the relation $P_i(P_i-1)=0$, i.e., $P_i^2=P_i$, we obtain
\begin{equation}
  F=-\sum_{i<j,\atop \hat{P}_i\hat{P}_j=0}P_iP_j-\sum_{\alpha}\left(2\sum_{i<j}P_{k_i^\alpha}P_{k_j^\alpha}-\sum_{i=1}^nP_{k_i^\alpha}+1\right).\label{eq:rayF}
\end{equation}
Third, the state-independent noncontextuality inequality reads $\langle F\rangle<-1$ and the quantum violation is given by $\langle F\rangle=0$ for all quantum states $\rho$. It is interesting to note that the inequality obtained by using the general approach is equivalent to the one in \cite{Yu.Tong2014}, which is given by trial and error.

For the case considered by Badzi\c{a}g \textit{et al} \cite{Badziag.etal2009}, where $\mathcal{S}$ includes $N$ bases, $\mathcal{S}_\alpha=\{P_{k_1^\alpha},P_{k_2^\alpha},\dots,P_{k_n^\alpha}\}$, $\alpha=1,2,\dots,N$, and every pair of orthogonal rays belongs to some  $\mathcal{S}_\alpha$, the \textit{complete} set of normalized polynomials can be taken as $\mathcal{P}_c=\{\sum_{i=1}^nP_{k_i^\alpha}-1=0|\alpha=1,2,\dots,N\}$. In this special case, equation (\ref{eq:rayF}) is reduced to
\begin{equation}
 F=-\sum_{\alpha}(2\sum_{i<j}P_{k_i^\alpha}P_{k_j^\alpha}-\sum_{i=1}^nP_{k_i^\alpha}+1),
  \label{eq:rayFs}
\end{equation}
and state-independent noncontextuality inequality reads
\begin{equation}
\sum_{\alpha=1}^N(\sum_{i=1}^n\langle P_{k_i^\alpha}\rangle-2\sum_{i<j}\langle P_{k_i^\alpha}P_{k_j^\alpha}\rangle)\le N-1,
\label{eq:rayB}
\end{equation}
which is valid for the NCHV model, but is violated by all quantum states $\rho$ since the expectation is $N$ in quantum mechanics.
In the previous work on noncontextuality inequalities, observables were usually chosen to be $\{-1,1\}$-dichotomic. If we let $A_i=1-2P_i$, the expressions (\ref{eq:rayF}),(\ref{eq:rayFs}) and (\ref{eq:rayB}) can be rewritten with observables $A_i$. For example, the inequality (\ref{eq:rayB}) becomes an equivalent form,
\begin{equation}
\sum_{\alpha=1}^N(2(n-2)\sum_{i=1}^n\langle A_{k_i^\alpha}\rangle-2\sum_{i<j}\langle A_{k_i^\alpha}A_{k_j^\alpha}\rangle)\le (n^2-3n+4)N-4,
\label{eq:rayC}
\end{equation}
which is valid for the NCHV model, but is violated by all quantum states $\rho$ since the expectation is $(n^2-3n+4)N$ in quantum mechanics. Compared with the noncontextuality inequality constructed by Badzi\c{a}g \textit{et al}, in which correlations of $n$ compatible observables are involved, the inequalities derived by our approach, which only involve correlations of two compatible observables, are simpler and should be more feasible experimentally.

\section{Application to proof of the KS theorem based on parity arguments}

Let $\mathcal{S}=\{A_1,A_2,\dots,A_\mu\}$ be a set of observables that provides a proof of the KS theorem based on parity arguments, i.e., a parity proof, in an $n$-dimensional Hilbert space. The observables in $\mathcal{S}$ are $\{-1,1\}$-dichotomic \footnote{For a parity proof based on rays, it can be convert to an equivalent form by replacing every ray $P_i$ by an $\{-1,1\}$-dichotomic observable $A_i=1-2P_i$.}, and there are $N$ measurement contexts in $\mathcal{S}$, denoted as $\mathcal{S}_\alpha=\{A_{k^\alpha_1},A_{k^\alpha_2},\dots,A_{k^\alpha_m}\}, ~\alpha=1,2,\dots,N$, which satisfy the expressions $\hat{A}_{k^\alpha_1}\hat{A}_{k^\alpha_2}\dots \hat{A}_{k^\alpha_m}=\delta_\alpha I_n$, where $\delta_\alpha$ equals to $1$ or $-1$ and $m=m(\alpha)$ is the number of observables in $\mathcal{S}_\alpha$. $\mathcal{S}$ being a parity proof means that these $N$ measurement contexts satisfy the relations: (i) $\delta_1\delta_2\cdots \delta_N=-1$, and (ii) each $A_i$ appears an even number of times in total in all $N$ measurement contexts, i.e., the number of measurement contexts that contain any $A_i$ is even. According to the expressions $\hat{A}_{k^\alpha_1}\hat{A}_{k^\alpha_2}\dots \hat{A}_{k^\alpha_m}=\delta_\alpha I_n$, we have the equations $v(A_{k^\alpha_1})v(A_{k^\alpha_2})\dots v(A_{k^\alpha_m})=\delta_\alpha$ if the value assignment $v$ satisfies the sum rule and product rule. However, these equations, for any value assignment, cannot be simultaneously valid, because there is a contradiction if they are used to calculate $\prod_{\alpha=1}^N v\left(A_{k^\alpha_1}A_{k^\alpha_2}\dots A_{k^\alpha_m}\right)$. On one hand, $\prod_{\alpha=1}^N v\left(A_{k^\alpha_1}A_{k^\alpha_2}\dots A_{k^\alpha_m}\right)=\prod_{\alpha=1}^N \delta_\alpha=-1$, but on the other hand, $\prod_{\alpha=1}^N v\left(A_{k^\alpha_1}A_{k^\alpha_2}\dots A_{k^\alpha_m}\right)=\prod_{\alpha=1}^N v(A_{k^\alpha_1})v(A_{k^\alpha_2})\dots v(A_{k^\alpha_m})=1$, since the number of each $A_i$ in the expression is even. This is the main idea of a proof of the KS theorem based on parity arguments.

We now derive a state-independent noncontextuality inequality from the parity proof.
The above discussion shows that $A_{k^\alpha_1}A_{k^\alpha_2}\dots A_{k^\alpha_m}-\delta_\alpha=0$ and that at least one of the equations, $v(A_{k^\alpha_1})v(A_{k^\alpha_2})\dots v(A_{k^\alpha_m})-\delta_\alpha=0$, is invalid for each value assignment $v$. Therefore, $\{A_{k^\alpha_1}A_{k^\alpha_2}\dots A_{k^\alpha_m}-\delta_\alpha~|~\alpha=1,2,\dots,N\}$ comprise a \textit{complete} set of polynomials. The normalized forms of the $N$ polynomials are $\{\frac{1}{2}(A_{k^\alpha_1}A_{k^\alpha_2}\dots A_{k^\alpha_m}-\delta_\alpha)=0~|~\alpha=1,2,\dots,N\}$, since $\delta_\alpha=\pm 1$ and $v(A_{k^\alpha_1})v(A_{k^\alpha_2})\dots v(A_{k^\alpha_m})=-\delta_i$ if $A_{k^\alpha_1}A_{k^\alpha_2}\dots A_{k^\alpha_m}\neq \delta_\alpha$. We then obtain the \textit{complete} set of normalized polynomials. Substituting them into equation (\ref{eq:defF}) and using the relations $(A_i-1)(A_i+1)=0$, i.e., $A_i^2=1$, we obtain
\begin{equation}
  F=-\frac{N}{2}+\frac{1}{2}\sum_{\alpha=1}^N\delta_\alpha \prod_{i=1}^{m(\alpha)} A_{k^\alpha_i}.
  \label{eq:parityF}
\end{equation}
The state-independent noncontextuality inequality can be written as
\begin{equation}
  \sum_{\alpha=1}^N\delta_\alpha\langle \prod_{i=1}^{m(\alpha)} A_{k^\alpha_i}\rangle\le N-2,
  \label{eq:Cabello}
\end{equation}
which is violated by all quantum states $\rho$ since $\sum_{\alpha=1}^N\delta_\alpha\langle\prod_{i=1}^{m(\alpha)} A_{k^\alpha_i}\rangle=N$ in quantum mechanics.

By applying our approach to proofs of the KS theorem based on parity arguments, we have obtained a general expression of the state-independent noncontextuality inequality. With this result, all known parity proofs, for example, those in Refs. \cite{Mermin1990,Peres1991,Pagonis.etal1991,Kernaghan1994,Cabello.etal1996,Cabello1999,Cabello2001,Planat2012,Waegell.Aravind2012,Waegell.Aravind2013,Ruuge2012,Lisonek.etal2014b} can be easily converted to state-independent noncontextuality inequalities, and the inequalities in Refs. \cite{Cabello2008} and \cite{Lisonek.etal2014a} are special cases of equation (\ref{eq:Cabello}).

\section{Conclusions}

We have shown that a proof of the KS theorem can always be converted to a state-independent noncontextuality inequality. This conclusion is true for all kinds of proofs of the KS theorem, based on rays or any other observables. It is interesting to note that our constructive proof actually provides a general approach of deriving a noncontextuality inequality from a proof of the KS theorem. By following the steps described by equations (\ref{PC}) to (\ref{eq:ineq}), one can derive a state-independent noncontextuality inequality from any proof of the KS theorem.  As examples, we have applied our approach to two kinds of well-known proofs of the KS theorem, i.e., proofs based on rays and proofs based on parity arguments. Certainly, our approach is applicable to all kinds of proofs of the KS theorem,including those in \cite{Kernaghan.Peres1995,Waegell.Aravind2012,Toh2013a,Toh2013b}, not limited to the two examples.

Compared with the methods proposed in \cite{Badziag.etal2009} and \cite{Yu.Tong2014}, which are applicable only to proofs of the KS theorem based on rays, the present approach is applicable to all kinds of proofs of the KS theorem. The inequalities obtained by our approach are simpler than the previous results in \cite{Badziag.etal2009} and as simple as the ones in \cite{Yu.Tong2014} when it is applied to proofs based on rays, and new state-independent noncontextuality inequalities are given when it is applied to proofs based on other observables.

\section*{Acknowledgments}
This work was supported by NSF China through Grant No. 11175105 and the National Basic Research Program of China through Grant No. 2015CB921004. D.M.T. acknowledges support from the Taishan Scholarship Project of Shandong Province.

\section*{References}

\providecommand{\newblock}{}

\end{document}